\documentstyle[12pt,amssymb]{article}
\input epsf
\def\theequation{\arabic{section}.\arabic{equation}}
\newcounter{subequation}[equation]
\makeatletter
\@addtoreset{equation}{section}
\expandafter\let\expandafter\reset@font\csname reset@font\endcsname
  {\endeqnarray\stepcounter{equation}}
\makeatother
\newcommand{\be}{\begin{equation}}
\newcommand{\ee}{\end{equation}}

\normalbaselines
\begin{document}
\title{A New Type of Conformal Dynamics}
\author{
P.C. Stichel\\
An der Krebskuhle 21\\ D-33619 Bielefeld, Germany \\
e-mail:pstichel@gmx.de
\\ \\
W.J. Zakrzewski
\\
Department of Mathematical Sciences, University of Durham, \\
Durham DH1 3LE, UK \\
 e-mail: W.J.Zakrzewski@durham.ac.uk}
\date{}
\maketitle

\begin{abstract}

 We consider the Lagrangian particle model introduced in [1] for zero
mass but nonvanishing second central charge of the planar Galilei group.
Extended by a magnetic vortex or a Coulomb potential the model exibits
conformal symmetry. In the former case we observe an additional $SO(2,1)$
hidden symmetry. By either a canonical transformation with constraints
or by freezing scale and special conformal transformations at $t=0$ we
reduce the six-dimensional phase-space to the physically required four
dimensions. Then we discuss bound states (bounded solutions) in quantum
dynamics (classical mechanics). We show that the Schr\"odinger equation
 for the pure
vortex case may be transformed into the Morse potential problem thus providing
us with
 an explanation of the hidden $SO(2,1)$ symmetry.

\end{abstract}

\section{Introduction}

In a recent paper [1] Lukierski and the present authors introduced a dynamical
realisation of the two-fold centrally extended planar Galilei group [2]
by means of the higher order particle Lagrangian
\be
L_0\,=\,\frac{m}{2}\dot x_i\sp2\,-\,\kappa\,\epsilon_{ij}\,\dot x_i\, \ddot x_j.
\ee

In (1.1) $m$ and $2\kappa$ are the two central charges of the group with
$2\kappa$ being given by the nonvanishing Poisson-bracket between the boost
generators
\be
\{ K_i,\,K_j\}\,=\,2\kappa \epsilon_{ij}.
\ee

The Lagrangian (1.1) leads to a six-dimensional phase space which, for
$m\ne 0$ may be split into a four-dimensional physical sector, described
by noncommuting variables, and an auxiliary two-dimensional sector [1,3].
Both sectors are dynamically independent from each other.

 The model (1.1) has been successfully extended by coupling  it to
either a scalar potential [1,3], electromagnetic and other gauge interactions
[4,5] or by its supersymmetrization [6].

Furthermore, field-theoretic models which allow for the nonvanishing of the second central charge of the planar Galilei group have also recently been studied [7,8]
by Horvathy, Martina and one of the present authors (PCS).

In the present paper we examine conformal symmetry of the model determined
by (1.1) to which some further interactions are added.

However, let us point out at the onset that even in the noninteracting
case we have a problem; namely, the two terms in (1.1)
have different transformation properties with respect to dilatations
and special conformal transformations. To see this we introduce
a function
\be 
f(t):=\, a\,+\,bt\,+\,ct\sp2
\ee
with infinitesimal parameters $a$, $b$ and $c$. Then the first term, i.e. 
$\dot x_i\sp2$ in (1.1) is quasi-invariant with respect to [9,10]
\be
\delta x_i\,=\,f\dot x_i\,-\,\frac{1}{2} \dot f x_i
\ee
while the second term ($\epsilon_{ij}\dot x_i\ddot x_j$) is 
quasi-invariant with respect to
\be
\delta x_i\,=\,f \dot x_i\,-\,\dot f x_i
\ee
i.e. $x_i$ has an anomalous scale dimension in this case.

If we keep only the first term in (1.1) we can easily extend our model to the well known models with either an inverse square potential [9] or a magnetic
vortex [10]. In order to construct particle models with a nonvanishing
second central charge of the planar Galilei group we must, therefore, put the 
first central charge, the mass $m$, equal to zero.

Our paper is organized as follows. In section 2 we show that we preserve 
conformal invariance if we add to (1.1) a Coulomb potential or a magnetic
vortex interaction. In the case 
of the pure vortex we find an additional hidden $SO(2,1)$ symmetry. As for $m=0$ the decomposition of the six-dimensional
phase-space into a physical and an auxiliary sector is not possible so in section 3 we perform a canonical transformation with constraints giving 
us the reduction to a reduced phase-space with the physically 
required four dimensions. Freezing a symmetry turns out to be an alternative 
approach to that. In Section 4 we look for bounded solutions of the classical equations of motion in this reduced phase-space.  Quantum dynamics
is discussed in section 5. We show that the Schr\"odinger equation for the pure vortex can be transformed into the Morse potential problem [11].
This explains the $SO(2,1)$ symmetry found in the classical case in the original phase space.

\section{An Exotic Particle Model with a Coulomb and a Magnetic Vortex
Interaction}

We consider here a massless particle 
with a nonvanishing second central charge of the Galilei group in two dimensions, whose motion is
goverened by the equations of motion which follow from the Lagrangian
\be 
L\,=\,L\sp{0}\,-\,V(\vec{x},\dot{\vec{x}}),
\ee
where \cite{one}
\be
L\sp{0}\,=\,
\frac{\theta}{2}\,\epsilon_{ij}
\,\dot x_i\,\ddot x_j
\ee
i.e. we have put $\kappa=-\frac{\theta}{2}$ in the notation of \cite{one}.

The potential $V$ consists of two terms (Coulomb + vortex)
i.e. it is given by
\be 
V\,=\,V_1\,+\,V_2,
\ee
where
\be
V_1\,=\,\frac{\lambda}{\vert\vec x\vert}
\ee
and
\be
V_2\,=\,-\frac{g}{\vert \vec x\vert\sp2}\,\epsilon_{ij}\dot x_ix_j.
\ee

Note that $V_2$ describes the potential of the magnetic vortex field
located at the origin [10] with flux $\phi=\frac{2\pi g}{e}$.


It is well known that $V_2$ may be written as a total time derivative
of a function which is singular at the origin. Therefore $V_2$
does not contribute to the classical equations of motion
in the punctured plane. However, in order for all the equations to be
valid in the whole plane, and to have a smooth transition to the quantum case,
in the rest of the paper we will leave $V_2$ in the form (2.5).

\subsection{Hamiltonian and Symplectic Structure}
\subsubsection{First-order formalism}
Here we use the first order formalism, which for $L_0$ 
was used in \cite{one}. 
We have
\be
L\,=\,P_i(\dot x_i-y_i)\,+\,\frac{\theta}{2}\epsilon_{ij}y_i\dot y_j\,-\,\frac{\lambda}{r}
\,+\,\frac{g\epsilon_{ij}y_ix_j}{r\sp2}\label{Lagr}
\ee
with $r:=\vert \vec x\vert$.

The equations of motion which follow from this $L$ are given by
\be
i)\quad \hbox{Var}\quad P_i\qquad \rightarrow\qquad \dot x_i\,=\,y_i \phantom{\left(\frac{\epsilon_{kl}y_kx_l}{r\sp2}\right)\,+\,\frac{\lambda x_i}{r\sp3}}
\ee
\be
ii)\quad \hbox{Var}\quad x_i\qquad \rightarrow\qquad\dot P_i\,=\,g\partial_{i}\left(\frac{\epsilon_{kl}y_kx_l}{r\sp2}\right)\,+\,\frac{\lambda x_i}{r\sp3}
\ee
\be
iii)\quad \hbox{Var}\quad y_i\qquad\rightarrow\qquad \dot y_i\,=\,-\frac{1}{\theta}\left(
\epsilon_{ij}P_j\,+\,\frac{gx_i}{r\sp2}\right).\phantom{aaa}
\ee
\subsubsection{Hamiltonian}
The Hamiltonian which follows from (\ref{Lagr}) is then given by
\be
H\,=\,P_iy_i\,+\,\frac{\lambda}{r}\,+\,g\frac{\epsilon_{ij}x_iy_j}{r\sp2}
\ee

\subsubsection{Symplectic structure}
From 
$\dot Y\,=\,\{Y,\,H\}$, where
$Y\in (x_i,\,y_i,\,P_i)$ 
we have
\be
\{x_i,\,P_k\}\,=\,\delta_{ik},
\ee
\be
\{y_i,\,y_k\}\,=\,-\frac{\epsilon_{ik}}{\theta}
\ee
with all other Poisson brackets vanishing.

Note that the interaction terms in (\ref{Lagr}) do not change the symplectic
structure based on $L_0$ (cp. \cite{one}).

\subsection{Symmetries}
\subsubsection{Conformal transformations}
Define the functions
\be
D:\,=\,tH\,-\,x_iP_i
\ee
and
\be
K:\,=\,-t\sp2H\,+\,2tD\,+\,2\theta\epsilon_{ij}x_iy_j.
\ee
Note that due to (2.7-9) both these functions are conserved, i.e.
\be
\frac{d}{dt}D\,=\,\frac{d}{dt}K\,=\,0.
\ee
Moreover, note also that $D$ and $K$, together with $H$, build the operators
of the conformal algebra, i.e. we have
\be
\{D,\,H\}\,=\,-H
\ee
\be
\{K,\,H\}\,=\,-2D
\ee
\be
\{D,\,K\}\,=\,K
\ee
as can be checked by using the Poisson brackets (2.11-12).

The Casimir of the conformal algebra (2.16-18)
is then given by
\be
C\,=\,HK\,-\,D\sp2\,=\,2\theta H \epsilon_{ij}x_iy_j\,-\,(x_iP_i)\sp2,
\ee
where the last equality follows from (2.13) and (2.14).

Observe that $D$ and $K$ are, respectively, the generators of dilatations
and of special conformal transformations. The general conformal
transformation of a phase-space element $Y$ is thus given by
\be
\delta Y\,=\,a\{Y,\,H\}\,+\,b\{Y,\,D\}\,+\,c\{Y,\,K\},\label{infi}
\ee
where $a$, $b$ and $c$ are infinitesimal numbers.

Remark: For $Y=x_i$ we have from (\ref{infi})
\be
\delta x_i\,=\,fy_i\,-\,\dot f x_i,\label{inf}
\ee
where (cp. (1.3))
\be
f(t)\,=\,a\,+\,bt\,+\,ct\sp2.
\ee

Note that (\ref{inf}) differs from the corresponding expression in a conventional massive theory with an inverse square potential
\be
\hat L\,=\,\frac{m}{2}\,\dot x_i\sp2\,-\,\frac{\lambda}{r\sp2}
\ee
where the symmetry transformation is given by [9]
\be
\hat \delta x_i\,=\,f\dot x_i\,-\,\frac{1}{2}\dot f x_i.
\ee

\subsubsection{Rotations}
Note that the conserved generator of planar rotations is given by
(cp. \cite{one})
\be
J\,=\,\epsilon_{ij}x_iP_j\,+\,\frac{\theta}{2}y_i\sp2.
\ee

\subsubsection{Hidden Symmetry of the pure vortex case}

Let us observe that when $\lambda=0$, i.e. when we have a pure vortex case, our equations (2.7-9) possess two additional conserved quantities, namely:
\be
\frac{2}{\theta}J_1\,=\,\cos\,2\varphi\left(\frac{(\epsilon_{ij}x_iy_j)\sp2}{r\sp2}\,-\,\frac{J+g}{\theta}\right)\,+\,\zeta\,\sin\,2\varphi
\ee
and
\be
\frac{2}{\theta}J_2\,=\,\sin\,2\varphi\left(\frac{(\epsilon_{ij}x_iy_j)\sp2}{r\sp2}\,-\,\frac{J+g}{\theta}\right)\,-\,\zeta\,\cos\,2\varphi,
\ee
where $r$, $\varphi$ are polar coordinates of the vector $\{x_i\}$ and $\zeta$
is defined by
\be
\zeta:=\,\frac{x_iy_i}{r\sp2}\,\epsilon_{kl}x_ky_l\,-\,\frac{1}{\theta}x_iP_i.
\ee

Let us define
\be
J_{\pm}:=\,J_1\,\pm\,iJ_2\quad\hbox{and}\quad J_0:=\,\frac{1}{2}(J+g).
\ee
Then, as is easy to check, these quantities satisfy a $SO(2,1)$ Poisson
bracket algebra
\be
\{J_0,\,J_{\pm}\}\,=\,\mp\,iJ_{\pm}
\ee
and
\be
\{J_+,\,J_-\}\,=\,2iJ_0,
\ee
where we have used
\be
\{\zeta,\,x_k\}\,=\,\frac{x_k}{\theta}
\ee
and
\be
\{\zeta,\,\epsilon_{ij}x_iy_j\}\,=\,0.
\ee

Surprisingly, the algebra's Casimir $J_o\sp2-J_+J_-$ is proportional 
to the Casimir $C$ of the conformal algebra given by (2.19), i.e.
\be
I:=\,J_0\sp2\,-\,J_+J_-\,=\,\frac{1}{4}C.
\ee

In deriving (2.34) we have used the planar vector identity
\be
y_i\,=\,x_i\,\frac{x_ky_k}{r\sp2}\,-\,\epsilon_{ij}x_j\,\frac{\epsilon_{kl}x_ky_l}{r\sp2}.
\ee

Each term in (2.26-27) has a vanishing Poisson bracket with $x_iP_i$, i.e.
$$ \{x_iP_i,\,\varphi\}\,=\,0
$$
\be
 \{x_iP_i,\,\frac{(\epsilon_{kl}x_ky_l)\sp2}{r\sp2}\}\,=\,0,\quad \hbox{and}\quad  \{x_iP_i,\,\zeta\}\,=\,0
\ee
leading to
\be
\{D,\,J_{\pm}\}\,=\,0.
\ee

Using (2.14) and (2.33) we see that $J_{\pm}$ are invariant also with respect
to the special conformal transformations, i.e.
\be
\{K,\,J_{\pm}\}\,=\,0.
\ee
As the conformal algebra (2.16-18) is isomorphic to the $SO(2,1)$ algebra [9]
we conclude that in the pure vortex case the total symmetry group $G$ is given
by
\be G\,=\,SO(2,1)\otimes SO(2,1).
\ee

\subsection{Classification of Solutions}

From (2.10) and (2.7-9) we see that
\be \frac{E}{\theta}\,=\,\frac{d\sp2}{dt\sp2}\,(\epsilon_{ij}\,x_i\,\dot x_j).
\ee
Integrating this we get
\be
\epsilon_{ij}\,x_i\,\dot x_j\,=\, \frac{E}{2\theta}t\sp2\,+\,A_1t\,+\,A_2,
\label{sol}
\ee
where
$A_{1,2}$ are constants. Let us note that, in polar coordinates, the left hand side of (\ref{sol})
is given by $r\sp2\dot\varphi$.

Next we substitute this expression, together with (2.9), into (2.19) and find that
\be
C\,=\,2E\theta A_2\,-\,\theta\sp2 A_1\sp2.
\ee

Thus we see that each solution of the equations of motion is characterised by the three constants: 
$E$, $A_1$ and $A_2$.

\section{Canonical Transformation}

Our phase space is six-dimensional. In order to reduce it to the physically
required four dimensions we consider, in the following, an appropriate 
canonical transformation with constraints.

Let us return to (\ref{sol})
and define
\be
h(t)\,:=\, \epsilon_{ij}\,x_i\,\dot x_j\, =\,\frac{E}{2\theta}t\sp2\,+\,A_1t\,+\,A_2.
\label{sol1}
\ee

Next we look for a transformation
\be
t\,\rightarrow\,t',\quad x_i(t)\,\rightarrow\,x_i'(t')
\ee
which brings (\ref{sol1})
to the form
\be
\epsilon_{ij}\,x'_i\,\dot x'_j\,=\,1.\label{condition}
\ee

This is achieved by putting
\be
\frac{dt'}{dt}\,=\,\frac{1}{h(t)}\qquad x_i(t)\,=\,h(t)\,x'_i(t').
\label{canh}
\ee
Integrating we find

\be
t'(t)\,=\,\left\{\begin{array}{c}{-\frac{2}{\sqrt{-\frac{C}{\theta\sp2}}}\,Arth\frac{A_1+\frac{E}{\theta}t}{\sqrt{-\frac{C}{\theta\sp2}}}},\quad C<0\\
-\frac{2}{A_1+\frac{E}{\theta}t},\quad\qquad C\,=\,0\\
\frac{2}{\sqrt{\frac{C}{\theta\sp2}}}\,arctg\,\frac{A_1+\frac{E}{\theta }t}{\sqrt{\frac{C}{\theta\sp2}}},\quad C\,>0,\end{array}\right.
\ee

Note that here $C=2E\theta A_2-\theta\sp2 A_1\sp2$.


Next we extend (3.4) to a canonical transformation:
\be
(\vec x, \,\vec y,\, \vec P)\,\rightarrow\, (\vec x',\,\vec y',\,\vec P')
\ee
by putting
\be
y'_i(t')\,=\,y_i(t)\,+\,f_1(t)\,x_i(t),
\ee
\be
P'_i(t')\,=\,f_2(t)\,P_i(t)\,+\,f_3(t)\,\epsilon_{ij}\,y_j(t)\,+\,f_4(t)\,\epsilon_{ij}\,x_j(t).
\ee
Using the Poisson bracket structure  (2.11)
\be \{x'_i,\,P'_j\}\,=\,\delta_{ij}
\ee
we find
\be 
f_2(t)\,=\,h(t).\ee
Moreover,
\begin{eqnarray}
\{y'_i,\,y'_j\}\,=\, -\frac{1}{\theta} \,\epsilon_{ij}\\
\{x'_i,\,y'_j\}\,=\,\{x'_i,\,x'_j\}\,=\,0
\end{eqnarray}
which are all automatically satisfied
and
\be
\{y'_i,\,P'_j\}\, =\, 0\quad \rightarrow\quad f_1h\,-\,\frac{1}{\theta}f_3\,=\,0
\ee
\be
\{P'_i,\,P'_j\}\,=\,0\,\quad \rightarrow \quad 2f_2f_4\,-\,\frac{1}{\theta}f_3\sp2\,=\,0
\ee
Furthermore, we require that, on shell, 

\be 
\dot x'_i\,=\,y'_i,
\ee
which gives us
\be
f_1\,=\,-\frac{\dot h}{h}\ee
while the other conditions are satisfied if we put
\be
f_3\,=\,-\theta \dot h
\ee
\be
f_4\,=\,\frac {\theta\dot h\sp2}{2h}.
\ee

Next we solve (3.7-8) with the functions given above. We find (all primed variables
are functions of $t'$)
\begin{eqnarray}
x_i\,=\,h\,x'_i,\quad y_i\,=\,y'_i\,+\,\dot h\,x'_i,\\
hP_i\,=\,P'_i\,+\, \theta \,\dot h\,\epsilon_{ij}\,y'_j\,+\,\frac{1}{2}\,\theta\,\dot h\sp2\,\epsilon_{ij}\,x'_j.
\end{eqnarray}

The new Lagrangian is then given by $L'=L\frac{dt}{dt'}$ and so takes the form
\be
L'\,=\,P'_i(\dot x'_i-y'_i)\,+\,\frac{\theta}{2}\,\epsilon_{ij}\,y'_i\dot y'_j\,-\,\frac{1}{2\theta}
C\,\epsilon_{ij}x'_i\,y'_j\,-\,\frac{\lambda}{r'}\,+\,g\frac{\epsilon_{ij}y_i'x_j'}{r'\sp2},
\ee 
where we have neglected the total time-derivative term
\be 
\frac{d}{dt'}\left(\frac{\theta}{2}\,\dot h\,\epsilon_{ij}\,x'_iy'_j\right).
\ee

The new total Hamiltonian is then given by
\be
H'\,=\,P'_iy'_i\,+\,\frac{\lambda}{r'}\,+\,\frac{C}{2\theta}\,\epsilon_{ij}x'_iy'_j\,+\,g\frac{\epsilon_{ij}x_i'y_j'}{r'\sp2}.
\ee

When compared to the original Lagrangian (2.6) we note that (3.21) and
so (3.23) contains an additional interaction term corresponding to 
the interaction with a uniform magnetic field of strength $-\frac{C}{e\theta}$.

It is easy to derive the equations of motion which follow from our (new) Lagrangian. They
are
\be
\theta \epsilon_{ij}\dot{\ddot{x'_j}}\,+\, \frac{C}{\theta}\epsilon_{ij}\dot x'_j\,=\,\frac{\lambda x'_i}{r\sp3},  \label{thet}
\ee
where $ \vec{x}'\in R_0\sp2:=R\sp2\setminus\{0\}.$

However, the solutions of these equations will not, in general, satisfy (\ref{condition}). In order to
overcome this problem we note that the transformation (3.19-20) depends, through the constants
$E$, $A_1$ and $A_2$, for
 the initial conditions for the old (unprimed) variables:
\begin{eqnarray}
h(0)\,=\,\epsilon_{ij}\,x_i(0)\,\dot x_j(0)\,=\,A_2,\\
\dot h(0)\, =\, \epsilon_{ij}\,x_i(0)\ddot x_j(0)\,=\,A_1,\\
E\,=\,\theta\,\epsilon_{ij}\,\dot x_i(0)\,\ddot x_j(0)\,+\, \frac{\lambda}{r(0)}.
\end{eqnarray}

Thus we need to transform these relations into the corresponding combinations  of initial conditions on the 
 new variables $x'_i(t')$ at $t'_0=t'(0).$

From (3.4) and (3.1) we get
\begin{eqnarray}
x_i(0)\,=\,A_2\,x'_i(t'_0),\\
\dot x_i(0)\,=\,A_1\, x'_i(t'_0)\,+\,\dot x'_i(t'_0),\\
\ddot x_i(0)\,=\,\frac{E}{\theta}\,x'_i(t'_0)\,+\,\frac{A_1}{A_2}\dot x'_i(t'_0)\,+\,\frac{1}{A_2}\ddot x'_i(t'_0).
\end{eqnarray}
Therefore, defining 
\be
h'(t'):\,=\,\epsilon_{ij}x'_i(t')\,\dot x'_j(t')
\ee we see that
\be 
h'(t'_0)\,=\,1,\label{aa}
\ee
\be
\dot h'(t'_0)\,=\,0\label{bb}
\ee
and using (3.29)
\be
\ddot h'(t_0')\,=\,0.
\ee

It is now easy to see that the solutions of the equations of motion, which satisfy the above given
initial conditions, automatically satisfy the subsidary condition $h'(t')=1.$

To see this we note that the equations of motion for
$x_i$ and $y_i$ take the form
\be
\dot x'_i\,=\,y'_i
\ee and
\be
P'_i\,=\,\theta\,\epsilon_{ij}\,\dot y'_j\,+\,\frac{C}{2\theta}\,\epsilon_{ij}x'_j\,+\,g\frac{\epsilon_{ij}x_j'}{r'\sp2},
\ee
and the `new' energy $E'$ may be rewritten as
\be E'\,=\,\theta \ddot h'(t')\,+\,\frac{C}{\theta}
h'(t').
\ee 
When we use this expression at $t'=t'_0$, together with (3.32,34), we get
\be
E'\theta \,=\,C
\ee while solving the remaining differential equation, with the initial
conditions (\ref{aa}-33),   gives us, as required
\be
h'(t')\,=\,1.
\ee

\subsection{Constraint Analysis and the Reduced Phase Space}

Our Lagrangian is given by
\be
L'\,=\,P_i(\dot x_i-y_i)\,+\,\frac{\theta}{2}\,\epsilon_{ij}\,y_i\,\dot y_j\,-\,\frac{\lambda}{r}\,-\,\frac{C}{2\theta}
\,\epsilon_{ij}\,x_iy_j\,-\,g\frac{\epsilon_{ij}x_iy_j}{r\sp2}
,\ee
where we have dropped all the primes. As the system is subject to a subsidiary
condition 
\be
\epsilon_{ij}\,x_i\,y_j\,=\,1,
\ee
which is a primary constraint,
we have to, for reasons of consistency, following Dirac, require also that
\be
\frac{d}{dt}\left(\epsilon_{ij}\,x_i\,y_j\right)\,=\,0
\ee
which, due to (3.35-36), gives us a secondary constraint
\be
x_i\,P_i\,=\,0.
\ee

 Our constraints (3.41,43) are solved by
\be
y_i\,=\,zx_i\,-\,\frac{1}{r\sp2}\,\epsilon_{ij}\,x_j\label{cons3}
\ee and
\be
P_i\,=\,u\,\epsilon_{ij}\,x_j.\label{cons4}\ee
Solving (\ref{cons3}) and (\ref{cons4}) for $z$ and $u$ we find
\be
z\,=\,\frac{y_ix_i}{r\sp2},\qquad u\,=\,\epsilon_{ik}\frac{P_ix_k}{r\sp2}.
\label{solu}\ee

So the reduced phase-space variables are  given by
$x_i$, $u$ and $z$.

\subsection{Dynamics in the reduced phase space}
Following Jackiw [12] we insert the solutions for the constraints (3.44,45) into $L'$ (modulo a 
total divergence term) and obtain
\begin{eqnarray}
L'\,=\,u\epsilon_{ij}\,\dot x_i\,x_j\,+\, u\,-\,\frac{\lambda}{r}\,-\,\frac{C}{2\theta}\,-\,\frac{g}{r\sp2}\nonumber\\
+\,\frac{\theta}{2}\left((z\sp2\,+\,
\frac{1}{r\sp4})\epsilon_{ij}\,x_i\dot x_j\,-\,\frac{2z}{r\sp2}(x_i\dot x_i)\right).\label{uu}
\end{eqnarray}

Next we perform the point transformation
\be 
(x_i,u,z)\,\rightarrow \,(r,\varphi,J,z),
\ee
where $r$ and $\varphi$ are the polar coordinates in the plane, and $J$ is initially given by
\be
J\,=\,\epsilon_{ij}\,x_i\,P_j\,+\,\frac{\theta}{2}\,y_i\sp2\label{ang}
\ee
and so becomes, due to (3.44-45)
\be
J\,=\,r\sp2(-u+\frac{\theta}{2}z\sp2)\,+\,\frac{\theta}{2r\sp2}.
\ee

Then the Lagrangian takes the form
\be
L'\,=\,J\dot \varphi\,-\,\frac{\lambda}{r}\,-\,\frac{C}{2\theta}\,-\,\frac{J+g}{r\sp2}\,+\,
\frac{\theta}{2}\left(z\sp2\,+\,\frac{1}{r\sp4}\right)\,-\,\theta z\frac{\dot r}{r}.\ee

To find the equations of motion we vary $J$ and
find
\be
\dot \varphi\,=\,\frac{1}{r\sp2}.
\ee

Next we vary $z$ and then $\varphi$ and obtain, respectively
\be
\dot r\,=\,rz
\ee
and
\be
\dot J\,=\,0.\ee

Finally, varying $r$ we get
\be
\dot z\,=\,-\frac{\lambda}{\theta r}\,-\,\frac{2(J+g)}{\theta r\sp2}\,+\,\frac{2}{r\sp4}.
\ee

\subsection{Hamiltonian, Canonical EoM and PB algebra}
The Hamiltonian becomes
\be
H'\,=\,\frac{J+g}{r\sp2}\,-\,\frac{\theta}{2}\left( z\sp2\,+\,\frac{1}{r\sp4}\right)\,+\,\frac{\lambda}{r}\,+\,\frac{C}{2\theta}.\ee

Hence the canonical EoM 
\be
\dot A\,=\,\{A,\,H'\}
\ee
become identical (when used for $A\in(r,\varphi,J,z)$)
to the EoM (3.52-55) if we require the following Poisson bracket structure
\be
\{\varphi,\,J\}\,=\,1,\qquad \{z,\,r\}\,=\,\frac{r}{\theta}.
\ee

\subsection{Phase-space reduction by freezing a symmetry}

Instead of considering the phase-space reduction by means of a canonical transformation
with constraints we  can take a different approach.

We start with the following description of the 6-dimensional phase space:
\be
\{A_i\}_{i=1}\sp4\, \oplus \,\{B_i\}_{i=1}\sp2,
\ee
where 
\be
A_i\,\in\,\{\frac{r}{\epsilon_{ij}x_iy_j},\varphi,J,\zeta\}
\ee
and
\be
B_i\,\in\,\{\epsilon_{ij}x_iy_j,\,x_kP_k\}.
\ee

The Poisson brackets of the decomposition (3.59) decouple from each other
\be
\{A_i,\,B_k\}\,=\,0,\quad i=1,..4,\,k=1,2
\ee
as can be seen from (2.33) and (2.36) respectively.

Furthermore, the subspace $\{B_i\}_{i=1}\sp2$ is invariant with respect to the scale
and special conformal transformations at $t=0$:
$$\{\epsilon_{ij}x_iy_j,\,D\}_{t=0}\,=\,-\epsilon_{ij}x_iy_j,\quad \{\epsilon_{ij}x_iy_j,\,K\}_{t=0}\,=\,0$$
\be
\{x_kP_k,\,D\}_{t=0}\,=\,0\quad \hbox{and}\quad \{x_kP_k,\,K\}_{t=0}\,=\,-2\theta\epsilon_{ij}x_iy_j.
\ee
Thus the choice of the constraints (3.41) and (3.43) is equivalent to the fixing of the components of the second part
$\{B_i\}_{i=1}\sp2$ in (3.59) in a particular way and may be understood as freezing the scale and the special conformal
transformation at $t=0$.

The remaining phase space $\{A_i\}_{i=1}\sp4$ is then identical to our reduced phase space $\{r,\varphi,J,z\}$ because,
as we can see from (2.28) and (3.46), we have
\be
\zeta\,=\,z\ee
on the constraint surface (3.41,43).

Note that in this procedure we do not obtain the additional interaction with a uniform magnetic field
(as in (3.21) and (3.23)).

\section{Solutions of the Classical Equations of Motion}

Next we consider the solutions of the equations of motion, ie the equations (3.52-55).
First, we note that these equations have two integrals of motion, namely
\be
E'\,=\,\frac{C}{2\theta}\,+\,\frac{J+g}{r\sp2}\,+\,\frac{\lambda}{r}
\,-\,\frac{\theta}{2}\left((\frac{\dot r}{r})\sp2\,+\,\frac{1}{r\sp4}\right),
\ee
and the angular momentum $J$.
%

By using (3.38) we may rewrite (4.1) as
\be
\frac{J+g}{\theta}\,=\,V(r)\,+\,\frac{\dot r\sp2}{2},\label{EE}
\ee
where
\be
V(r):\,=\,\frac{1}{2r\sp2}\,-\,\frac{\lambda}{\theta}r\,+\,\frac{C}{2\theta\sp2}r\sp2.\ee

 We see that we have reduced the problem to that of a motion of a particle of ``mass" =1 moving in the potential
$W(r)= \frac{\lambda}{\theta}r\,+\,\frac{C}{2\theta\sp2}r\sp2$ with 
``angular momentum" =1 and ``energy" = $\frac{J+g}{\theta}$.

Note, that in this way our ``nonstandard" problem has become a ``standard" one. 

We have to consider three separate cases: $C=0$, $C>0$ and $C<0$.

\subsection{The case of $C=0$}
Thus
\be
V(r):\,=\,\frac{1}{2r\sp2}\,-\,\frac{\lambda}{\theta}r.
\ee
Let us concentrate our attention on the case of
\be
\frac{\lambda}{\theta}\,<\,0
\ee
i.e. of an attractive potential.

Then, we note that $\frac{J+g}{\theta}>0$. At the same time $V(r)$ has a minimum
at $r_0=\left(-\frac{\theta}{\lambda}\right)\sp{\frac{1}{3}}$ which,
given that $r_0>0$ tells us that $V(r_0)\,=\,\frac{3}{2}\left(\frac{\lambda}{\theta}\right)\sp{\frac{2}{3}}.$
Thus we have
\be
\frac{J+g}{\theta}\,\ge\,\frac{3}{2}\left(\frac{\lambda}{\theta}\right)\sp{\frac{2}{3}}.
\ee
Note that when 
\be
\frac{J+g}{\theta}\,=\,\frac{3}{2}\left(\frac{\lambda}{\theta}\right)\sp{\frac{2}{3}}
\ee
we have
\be
r(t)\,=\,r_0
\ee
i.e. we have a steady motion along a circle of radius $r_0$.

For $\frac{J+g}{\theta}> \frac{3}{2}\left(\frac{\theta}{\lambda}\right)\sp{\frac{2}{3}}$ 
we have a bounded motion. To see this we note that the integration of 
(\ref{EE}) gives us
\be
t\,=\,\pm\,\int_{r(0)}\sp{r(t)}\,dr'\,\left(2\left(\frac{J+g}{\theta}\,-\,V(r')\right)\right)\sp{-\frac{1}{2}}.
\ee

When $\frac{J+g}{\theta}=V(r)$ has at least 2 positive roots there is a region for the bounded motion
\be
0\,<r_{min}\,\le \,r\,\le \,r_{max}.
\ee

\subsection{The case of $C>0$}
When $C>0$ we have
\be
\frac{J+g}{\theta}\,=\,\frac{\dot r\sp2}{2}\,+\,\frac{1}{2r\sp2}\,-\,\frac{\lambda r}{\theta}\,+\,\frac{C}{2\theta\sp2}r\sp2
\ee
and so it is clear that the motion is bounded.

Thus we are left with having to discuss the $C<0$ case.
\subsection{The case of $C<0$}

Now we have
\be
\dot r\sp2\,=\,\frac{1}{r\sp2}\left[\frac{2(J+g)}{\theta}r\sp2\,-\,1\,+\,\frac{2\lambda}{\theta}r\sp3\,+\,\frac{B}{\theta\sp2}r\sp4\right]\,
\ee
where 
$B=\vert C\vert.$

Next we define
\be
F(r)\,=\,\frac{2(J+g)}{\theta}r\sp2\,-\,1\, +\frac{2\lambda}{\theta}r\sp3\,+\,
\frac{B}{\theta\sp2}r\sp4
\ee
which we rewrite as
\be
F(r)\,=\,\frac{B}{\theta\sp2}\left(r\sp4\,+\,\alpha r\sp3\,+\,\beta r\sp2\,+\,\gamma\right),
\ee
where $\alpha=\frac{2 \lambda\theta}{B}$, $\beta=\frac{2(J+g)\theta}{B}$
and $\gamma=-\frac{\theta\sp2}{B}$.

Then all the properties of the solution depend on the values
of the parameters $\alpha$, $\beta$ and $\gamma$.

 Clearly $F(r)\rightarrow\infty$ as $r\rightarrow\infty$.
Moreover $F(0)=\gamma<0$. As $r=0$ is an extremum of $F(r)$, in order
to have a bounded motion we need to have $F(r_1)>0$ and $F(r_2)<0$
where $r_{1}$ and $r_2$ are two further extrema of $F$, which 
should lie at $r>0$ with $r_1<r_2$.

Then the bounded motion would involve $r$ changing between two roots
of $F$ lying between $r=0$ and $r_1$, and, $r_1$ and $r_2$
respectively.

So we look at the extrema of $F$. Clearly
$r_1$ and $r_2$ are given by
\be
r_i\,=\,\frac{-3\alpha\pm\sqrt{9\alpha\sp2-32\beta}}{8},
\ee
with the lower (upper) sign for $r_1(r_2)$.

As both $r_i$ have to be positive
we require that (note that due to (4.5) $\alpha<0$)

\be 
9\alpha\sp2> 32\beta>0\quad \rightarrow\quad \lambda\sp2>\frac{16(J+g)B}{9\theta}, \quad J+g>0.
\ee

It is easy to show that at $r=r_i$ \ \ \   $F$ takes the values
\be
F(r_i)\,=\,\left(\left[-\frac{1}{2}\alpha\beta\,+\,\frac{9}{64}\alpha\sp3\right]r_i\,+\,
\gamma\,-\,\frac{\beta\sp2}{4}\,+\,\frac{3}{32}\alpha\sp2\beta\right)\frac{B}{\theta\sp2}.
\ee

Now we want to show that there is a nontrivial region of the parameters
$\alpha$, $\beta$ and $\gamma$ such that $F(r_1)>0$ while
$F(r_2)<0$.

For this purpose we take, in accordance with (4.16), a particular value of
$\beta$, namely
\be
\beta_0\,=\,\frac{3}{4}\,\frac{9\alpha\sp2}{32}.
\ee

For this value we have 
\be r_{i}\,=\,\frac{3}{16}\vert \alpha\vert(2\mp 1)
\ee
 and so
\be
F(r_i)\,=\,\left(\frac{27}{8\sp4}\,\alpha\sp4(\pm 1\,+\,1)\,-\,\frac{27\sp2}{2\times 8\sp5}\alpha\sp4\,+\,\gamma\right)\frac{B}{\theta\sp2}
\ee
with the upper (lower) sign for $r_1(r_2)$.

Then, obviously 
\be
F(r_2)\,<\,0
\ee
and also we have $F(r_1)>0$ iff
\be
\left( 2-\frac{27}{16}\right)\,\frac{27}{8\sp4}\,\alpha\sp4\,+\,\gamma\,>\,0
\ee
or equivalently
\be
\left\vert \frac{\lambda}{\theta}\right\vert\,>\,2.347\left(\frac{B}{\theta\sp2}\right)\sp{\frac{3}{2}}.
\ee

Due to the continuity of $F(r_i)$, as a function of $\beta$ we will have
(4.21-22), and therefore bounded motion, for a range of $\beta$ around $\beta_0$
if we choose the numerical factor in (4.23) as an appropriate function 
of $\beta$.

Thus we see that for $C<0$  we can have solutions that describe 
bounded motion too.

\subsection{Hidden Symmetry of the pure vortex case}

The canonical transformation (3.19-20) supplemented by the
constraints (3.41,43) leads to the following transformation rules
$$ \varphi\,\rightarrow\,\varphi,\qquad \frac{(\epsilon_{ij}x_iy_j)\sp2}{r\sp2}\,\rightarrow\,\frac{1}{r\sp2}$$
\be \hbox{and}\qquad \zeta\,\rightarrow\,z.
\ee

Thus, the conserved quantities $J_{1,2}$ given by (2.26-27) in the original phase-space
become
\be
\frac{2J_1}{\theta}\,=\,\cos\,2\varphi\,\left(\frac{1}{r\sp2}\,-\,\frac{J+g}{\theta}\right)\, +\, z\,\sin\,2\varphi\label{newone}
\ee
and
\be
\frac{2J_2}{\theta}\,=\,\sin\,2\varphi\,\left(\frac{1}{r\sp2}\,-\,\frac{J+g}{\theta}\right)\, -\, z\,\cos\,
2\varphi\label{newtwo}
\ee
in the reduced phase space. It is self-evident that, together with $J_0$, they still satisfy the
$SO(2,1)$ Poisson bracket algebra (2.30-31). The algebra's Casimir (2.34) is now given by
\be
I\,=\,\frac{\theta}{2}H,
\ee
where we have defined $H:=H'-\frac{C}{2\theta}$.

For the classical orbit we have from (4.25-26)
\be
\frac{2}{\theta}\,(J_1\,\cos\,2\varphi\,+\,J_2\,\sin\,2\varphi)\,=\,\frac{1}{r\sp2}\,-\,\frac{J+g}{\theta}
\ee
thus showing that the classical trajectory is then given by
\be
r(\varphi)\,=\,\left(\frac{2J_1}{\theta}\,\cos\,2\varphi\,+
\,\frac{2J_2}{\theta}\,\sin\,2\varphi\,+\,\frac{J+g}{\theta}\right)\sp{-\frac{1}{2}}.
\ee

An interesting case then arises when
\be
J_1\,=\,J_2\,=0
\ee
realised by means of the initial conditions
\be
z(t=0)\,=\,0,\quad r(t=0)\,=\,\left(\frac{J+g}{\theta}\right)\sp{-\frac{1}{2}}.
\ee

Then, due to $J_{1,2}$ being conserved, (4.31) hold at an arbitrary time $t$.
In this particular case the EoM (3.52-53) reduce to
\be
\dot{ \varphi}\,=\,\frac{1}{r\sp2},\quad \dot{r}\,=\,0,
\ee
which is equivalent to
\be
\dot{x_i}\,=\,-\frac{\epsilon_{ij}x_j}{r\sp2}.
\ee

Note that (4.33) describes the relative motion of two fluid vortices (see [13] and the literature
cited therein). 
Note also that (4.33) may be derived as a canonical EoM from any Hamiltonian of the form
\be
H\,=\,H(r)
\ee
if we have the symplectic structure
\be
\{x_i,\,x_j\}\,=\,-\frac{\epsilon_{ij}}{rH'(r)}.
\ee
In our case we would have
\be
H(r)\,=\,\frac{\theta}{2r\sp4}
\ee
derived from (3.51)$\vert_{\lambda=0}$ by means of the primary constraint $z=0$.

\section{Quantum Dynamics}
\subsection{Schr\"odinger equation in reduced phase space}
We return to our Hamiltonian (3.56) and proceed to its quantization.
Thus our Poisson brackets (3.58) now become the commutation 
relations ($\hbar=1$)
\be
[z,\,r]\,=\,i\,\frac{r}{\theta},\quad\quad [J,\,\varphi]\,=\,\frac{1}{i}.
\ee

Representing $J$ and $z$ by symmetric differential operators 
\be
J\,=\,\frac{1}{i}\partial_\varphi\,\quad\quad z\,=\,\frac{i}{2\theta}\,(r\partial_r\,+\,\partial_rr)\,=\,\frac{i}{\theta}(r\partial_r\,+\,\frac{1}{2}).
\ee
gives us the radial Schr\"odinger equation
\be
\left(\frac{1}{2\theta}\left(r\partial_r\,+\,\frac{1}{2}\right)\sp2\,+\,\frac{\lambda}{r}\,+\,\frac{\bar m}{r\sp2}\,-\,\frac{\theta}{2r\sp4}\,-\,E\right)
\varphi_{E,m}(r)\,=\,0\label{ea}
\ee
with $\bar m:=m+g,$ $m\in Z$ and $E:=E'-\frac{C}{2\theta}$.

\subsection{Asymptotic behaviour of bound state solutions}
\subsubsection{$r\rightarrow 0$}

The dominant contribution of (\ref{ea}), in this limit, comes from
\be
\left(\left(r\partial_r\,+\,\frac{1}{2}\right)\sp2\,-\,\frac{\theta\sp2}{r\sp4}\right)
\varphi_{E,m}(r)\,=\,0.\label{eb}
\ee
We make the ansatz \cite{Dong}
\be
\varphi(r)\,\,=\,r\sp{-b}\,e\sp{-\frac{a}{r\sp2}}
\ee
and find that 
\be
a\,=\,\frac{\vert \theta\vert}{2}
\ee
while $b$ is undetermined.
\subsubsection{$r\rightarrow \infty$}
In this limit the dominant part of (\ref{ea}) comes from
\be
\left(\left(r\partial_r\,+\,\frac{1}{2}\right)\sp2\,-\,2\theta E\right)
\varphi_{E,m}(r)\,=\,0,\label{eb}
\ee
which shows that, as $r\rightarrow \infty$,
\be
\varphi(r)\,\sim\, r\sp{-(\frac{1}{2}+\sqrt{2\theta E})}.
\ee

For $\varphi$ to be normalisable, i.e. to satisfy
$$\int_0\sp{\infty}\,drr\,\vert \varphi\vert\sp2\,<\,\infty $$
we have to require that
\be
E\theta\,>\,0,\quad \hbox{and}\quad \sqrt{2E\theta}\,>\,\frac{1}{2}.
\ee

Let us assume that $\theta>0$ from now on. The general ansatz for $\varphi$, respecting both limits (5.5, 5.8)
is then given by (cp. \cite{Dong})
\be
\varphi_{E,m}(r)\,=\,e\sp{-\frac{\theta}{2r\sp2}}\,F(r)\label{ee}
\ee
with
\be
F(r)\,\stackrel{\textstyle{\sim}}{\scriptscriptstyle{r\rightarrow\infty}}\,r\sp{-b},\quad b:=\,\,\frac{1}{2}\,+\,\sqrt{2\theta E}
\ee
and
\be
r\sp{N}F(r)\,\stackrel{\textstyle{\sim}}{\scriptscriptstyle{r\rightarrow 0}}\,o(1)
\ee
for $N$ larger than some $N_0\in N$.

Putting (\ref{ee}) into (\ref{ea}) we end up with a differential equation for $F$, which is
\be
\frac{1}{2\theta}\left( r\sp2\ddot F\,+\,2\left(\frac{\theta}{r}\,+\,r\right)\dot F\right)\,+\,
\left(\frac{\bar m-\frac{1}{2}}{r\sp2}\,+\,\frac{\lambda}{r}\,-\,E\,+\,\frac{1}{8\theta}\right)F\,=\,0.\label{eg}
\ee

We have not succeeded in solving this equation in full generality. In the next section
we make some comments on its solutions obtained by a power series
expansion.

\subsection{Some comments about the solutions of (5.13)}

First we perform a transformation
\be
r\,\rightarrow\,u:=\,\frac{1}{r},\quad\quad G(u)\,:=\,F(r)
\ee
and find that (\ref{eg}) has now become
\be
u\sp2\ddot G\,-\,2\theta u\sp3\dot G\,+\,\mu u\sp2G\,+\,(\nu u\,+\,\omega)G\,=\,0\label{eh}
\ee
where
\be
\mu:=2\theta(\bar m-\frac{1}{2}),\quad \nu:=\,2\theta \lambda,\quad \omega:=\,\frac{1}{4}\,-\,2\theta E.
\ee

Next we attempt to solve (\ref{eh}) by the generalised power series around $u=0$. Hence
we put
\be
G(u)\,=\,\sum_{n=0}\sp{\infty}\,a_n\,u\sp{n+\alpha}.\label{ek}
\ee
The indicial equation is then
\be
\alpha(\alpha-1)\,+\,\omega\,=\,0
\ee
with solutions
\be
\alpha_{\pm}\,=\,\frac{1}{2}\,\pm\,\sqrt{2\theta E}.\label{ej}
\ee

Note that $\alpha_+=b$.
Next we find that
\be
a_1\,=\,-\frac{\nu}{2\alpha}a_0\label{eaa}
\ee
and that, for the general $a_n$ ($n\ge2$) we have the three point recurrence relation
\be
((\alpha +n)(\alpha+n-1)+\omega)a_n\,+\,\nu a_{n-1}\,+\,(\mu-2\theta(\alpha+n-2))a_{n-2}\,=\,0.\label{ei}
\ee

Clearly, starting with $a_0\ne 0$, we can determine $a_1$ and then, using (\ref{ei}), successively,
all $a_k$. This is true for either choice of $\alpha$ in (\ref{ej}). Both series converge as
(\ref{ei}) shows that as $n\rightarrow \infty$ we have
\be
\frac{a_n}{a_{n-1}}\,\rightarrow\,\pm\sqrt{\frac{2\theta}{n}}.
\ee

We finish this section with a few comments
\begin{itemize}
\item
Note that the solutions (\ref{ek}) corresponding to $\alpha_-$ are non-normalisable (ie they describe
scattering states).
\item
For $\nu\ne 0$ the series (\ref{ek}) cannot terminate for a generic value of $\nu$. Hence we have 
no discrete spectrum.

To show this we assume the contrary; i.e. $a_n=0$ for $n\ge p+1$ and $a_p\ne 0$
Then from (\ref{ei}) (with $n=p+2$) we have $\mu-2\theta(b+p)=0$. Inserting this into
(\ref{ei}) for $n=p+1$ we find that
\be
a_{p-1}\,=\,-\frac{\nu}{2\theta}a_p\label{ebb}
\ee
and so we see that we have a linear system of $(p+1)$ equations ((\ref{eaa}), (\ref{ebb}) and (\ref{ei}) for $2\le n\le p$)
for ($p+1$) unknowns $\{a_n\}_{n=0}\sp{p}$. This system has nontrivial solutions if and only if the corresponding
determinant vanishes. But this is not the case for a generic value of $\nu$. 

Example. Take $p=1$. Then the system of two equations (\ref{eaa}, \ref{ebb}) has the determinant
$\hbox{det}\,=\,1-\frac{\nu\sp2}{4\alpha \theta}\,\ne\,0$.

Thus we see that the bound states (belonging to $\alpha_+$ (5.19) have a continuous spectrum, which is characteristic
of singular potentials (for more details see \cite{Frank}).
\item 
For $\nu=0$, (i.e. $\lambda=0$) we can show that $p$=even, $a_n=0$ for odd $n$ and the spectrum is discrete
as our system reduces to $\frac{p}{2}$ equations for $\frac{p}{2}+1$ unknowns.
\end{itemize}

\subsection{Hidden symmetry of the pure vortex problem}

Let us now return to the case of the pure vortex. In the classical case we had two conserved quantities
(4.25) and (4.26).

The same is true in the quantum case. We have
\be
J_+:=\,\theta\frac{e\sp{2i\varphi}}{2}\left(\frac{r\partial_r\,+\,\frac{1}{2}}{\theta}\,+\,\frac{1}{r\sp2}\,
-\,\frac{J+g+1}{\theta}\right)\label{newa}
\ee
and
\be
J_-:=J_+\sp{\dagger}\,=\,\theta\frac{e\sp{-2i\varphi}}{2}\left(-\frac{r\partial_r\,+\,\frac{1}{2}}{\theta}\,+\,\frac{1}{r\sp2}\,
-\,\frac{J+g-1}{\theta}\right).\label{newb}
\ee

Then,
as can be seen by straightforward calculation, 
 we get, as in the classical case, the $SO(2,1)$ algebra 
$$[J_0,\,J_{\pm}]\,=\,\pm J_{\pm}$$
\be
[J_+,\,J_- ]\,=\,-2 J_{0}.\label{mmm}
\ee

Their Casimir is now
\be
I:=\,J_0(J_0\,\mp\,1)\,-\,J_{\pm}J_{\mp}\,=\,\frac{\theta}{2}H\,-\,\frac{1}{4}.
\ee

Note that both $J_{\pm}$ and $I$ contain additional quantum correction terms
when compared to their classical values.

A realization of the $SO(2,1)$ algebra (5.26) in terms of differential operators
$J_{\pm},\,J_0$ which commute with $H$, (potential algebra realization) has
been considered already some time ago [16]. Such a realization allows only for 
the existence of the $D\sp{+}$ discrete
principal series of $SO(2,1)$ for $\theta>0$. 
 Then the discrete spectrum of $H$ can
 be obtained by a standard procedure (cp. [16]):

We start with the ground state of the angular momentum ladder
\be
J_-\,\Psi_{0,m}\,=\,0,
\ee
where, in the coordinate representation,
\be
\Psi_{0,m}\,=\,e\sp{im\varphi}\varphi_{E,m}(r).
\ee

Then (5.28) is solved by
\be
\varphi_{E,m}(r)\,=\,r\sp{-(\bar m-\frac{1}{2})}\,e\sp{-\frac{\theta}{2r\sp2}},
\ee
where, due to (5.8,9), resp. (5.26) and (5.27), we have
\be
E\,=\,E_{0,m}:=\,\frac{1}{2\theta}(\bar m-1)\sp2
\ee
with
$m=m_0,m_0+2,m_0+4$; $m_0:=\{\frac{3}{2}-g\},$
where $\{.\}$ denotes the $Z$-part.

We note an infinite-fold degeneracy of the energy spectrum, because the whole
angular momentum ladder
\be
\Psi_{p,m+2p}:=\,(K_+)\sp{p}\,\Psi_{0,m}.\quad p\in N
\ee
possesses the same energy (5.31).

Equivalently, at fixed angular momentum $m$, i.e. at fixed potential
in the Schr\"odinger eq. (5.3), we have a finite number of bound states with 
energy
\be
E_{p,m}\,=\,\frac{1}{2\theta}(\bar m\,-\,2p\,-\,1)\sp2, \qquad p=0,1,2,..\{\frac{m-m_0}{2}\}.
\ee

\subsection{Relation between the vortex problem and the Morse potential}

\subsubsection{Transformation of the Schr\"odinger equation into the Morse potential problem}

Let us perform the following change of variables
\be
r\,\rightarrow\,\rho:=\,2\,ln\,r
\ee
and
\be
\varphi_{E,m}(r)\,=\,e\sp{-\frac{\rho}{2}}\Phi_{E,m}(\rho).
\ee

Then our Schr\"odinger equation (5.3) transforms, for $\lambda=0$, into
\be
\left(\frac{2}{\theta}\partial_{\rho}\sp2\,+\,\bar m e\sp{-\rho}\,-\,\frac{\theta}{2}
e\sp{-2\rho}\,-\,E\right)\Phi_{E,m}(\rho)\,=\,0.\ee

Define
\be
e\sp{\rho_0}:=\,\frac{\theta}{\bar m}
\ee
(we have chosen $\theta>0$ and have $\bar m>0$).

Then our equation becomes
\be
\left( -\frac{2}{\theta}\partial_{\rho}\sp2\, +\, \frac{\bar m\sp2}{2\theta}\left(1\,-\,e\sp{-(\rho-\rho_0)}\right)\sp2
-\frac{\bar m\sp2}{2\theta}\,+\,E\right)\,\Phi_{E,m}\,=\,0\quad -\infty<\rho<\infty.
\ee
This is the Schr\"odinger equation for the Morse potential \cite{Morse} with `mass'=$\frac{1}{4}\theta$ and
$E\sp{Morse}\,=\,-E\,+\,\frac{\bar m\sp2}{2\theta}$.
For the spectrum one has [11] 
\be
E\sp{Morse}\,=\,\frac{2\bar m}{\theta}\left( n\,+\,\frac{1}{2}\right)\,-\,\frac{2}{\theta}\left(n\,+\,
\frac{1}{2}\right)\sp2,\quad n\in N
\ee
which corresponds to (cp. (5.37))
\be
E_{n,m}\,=\,\frac{1}{2\theta}\,\left(\bar m-1-2n\right)\sp2.
\ee

\subsubsection{Generalised hidden symmetry}

The operators $I_{\pm}$ (see (\ref{newa}-\ref{newb}) shift the angular momentum, at fixed energy, by $\pm$ two units.

We may ask what are the operators which shift energy at fixed angular momentum. As, at fixed $\bar m$, the energies
are not equidistant these shift operators are not elements of a linear algebra. However, a realisation in terms of
a quadratic $SU(2)$-algebra is possible \cite{ten} for the Morse potential. This could then be translated
to our case and would lead to the generalised form of the hidden symmetry 
(cp. also \cite{fourteen}).

 An explicit construction of such shift operators, as functions of the parameters
of the Schr\"odinger equation, is given in \cite{twelve}.

\section{Concluding Remarks}

The Lagrangian for a massive point particle with either an inverse
square potential in arbitrary dimensions $D$ [9], a magnetic monopole in
$D=3$ [20] or a point magnetic vortex in $D=2$ [10] exhibits conformal
symmetry corresponding to the scale dimension $-\frac{1}{2}$ of $x_i$
 (see (2.24)). But
in the present exotic model (2.1) the potential is of the Coulomb type
and $x_i$ has therefore anomalous scale dimension -1 (see (2.21)). We can
still add a point magnetic vortex because the vortex allows an arbitrary scale
dimension of $x_i$ [10]. An additional hidden $SO(2,1)$ symmetry is observed
in the pure vortex case.

The reduction of six-dimensional phase-space to the physically required
four dimensions is performed in an unconventional way. We can use either a
canonical transformation with constraints or a frozen symmetry at $t=0$.
Related to the frozen symmetry  method, but different in detail, is the
``geometric symmetry-breaking mechanism" introduced some time ago by Cho
[21]. We believe that both our methods are of a general importance.

The solutions of the classical EOM in the reduced phase-space show for $C<0$
a bounded motion for strong Coulomb coupling (see (4.23)). For the quantum
analogue one would expect bound states embedded in the continuum. But it
has been claimed [22]  that these bounded solutions, which are due to
potential barriers, would appear as scattering resonances in  quantum
dynamics.

In the quantum case we find a continuous spectrum of bound states for
the generic values of the Coulomb coupling $\lambda$. This is due to the
singular nature of the effective potential in (5.3). However, for $\lambda=0$
we observe the hidden $SO(2,1)$ symmetry leading to the well known
discrete spectrum of the equivalent Morse potential problem [11].

Further investigations of the present model are possible. In particular
a supersymmetric extension is called for.

\vskip 1cm
{\bf Acknowledgement:}
We are grateful to R. Jackiw for valuable suggestions, to P. Horvathy 
and J. Lukierski for 
comments and correspondence.

\textheight 8.8in \textwidth 6in

\end{document}